\documentstyle[12pt]{article}

\def\Journal#1#2#3#4{{#1}{\bf #2}, #3 (#4)}
\def\NCA{Nuovo Cimento}

\def\NPB{{Nucl. Phys.} B}
\def\NPA{{Nucl. Phys.} A}
\def\PLB{{Phys. Lett.}  B}
\def\PRL{Phys. Rev. Lett.}
\def\PRD{{Phys. Rev.} D}
\def\ZPC{{Zeits.f. Phys.} C}
\def\ZPA{{Zeits.f. Phys.} A}

\def\MPLA{{Mod. Phys. Lett.} A}

\begin{document}

\author{Anatoly Adamov \\
\vspace{-2 mm}
{\small \it Department of Physics}\\
\vspace{-2 mm}
{\small \it Tufts University}\\
{\small \it Medford, MA 02155 USA} \\
and \\
Gary R. Goldstein\footnote{email: ggoldste@tufts.edu} \\
\vspace{-2 mm}
{\small \it Center for Theoretical Physics} \\
\vspace{-2 mm}
{\small \it Laboratory for Nuclear Science} \\
\vspace{-2 mm}
{\small \it and Department of Physics} \\
\vspace{-2 mm}
{\small \it Massachusetts Institute of Technology} \\
\vspace{-2 mm}
{\small \it Cambridge, MA 02139 USA} \\
\vspace{-2 mm}
{\small \it and} \\
\vspace{-2 mm}
{\small \it Department of Physics}\footnote{Permanent address} \\
\vspace{-2 mm}
{\small \it Tufts University} \\
\vspace{-2 mm}
{\small \it Medford, MA 02155 USA}}

\title{\Large Excited State Contributions to the Heavy Baryon
Fragmentation Functions in a Quark-Diquark Model\thanks{This work is
supported in part by funds provided by the U.S. Department of
Energy (D.O.E.) \#DE-FG02-92ER40702 and \#DF-FC02-94ER40818.}}
\vspace{-0.5 in}
\date{{\small (MIT-CTP: \#3025 \hfill Submitted to: {\it Phys. Rev. D} 
\hfill 20 September 2000)}}
\vspace{-0.5 in}
\maketitle

\vspace{-0.5 in}
\begin{abstract}
Spin dependent fragmentation functions for heavy flavor quarks to fragment 
into heavy baryons are calculated in a quark-diquark model. The production
of intermediate spin 1/2 and 3/2 excited states is explicity included. The
resulting $\Lambda_b$ production rate and polarization at LEP energies are
in agreement with experiment. The $\Lambda_c$ and $\Xi_c$ functions are
also obtained. The spin independent $f_1(z)$ is compared to data. The  
integrated values for production rates agree with the data.
\end{abstract}

\newpage

\section{Introduction.}

Fragmentation of quarks and gluons into hadrons is an important subject
for which QCD, in both the soft and hard regions, should be
applicable. Because of the entanglement of both of these regions, it is
difficult to calculate fragmentation functions $in \ general$. However,
when fragmentation involves heavy flavor quarks and hadrons the situation
is clearer -- factorization of the soft and hard parts can obtain. Heavy
quark physics lends itself to perturbative QCD. The calculations are  
considerably simplified by the peculiar feature
of heavy physics just as in atomic physics, where
the presence of the heavy nucleus, with a mass much larger than the average
momentum transfer inside of the system, effectively reduces the number of
the degrees of freedom. A heavy quark or a heavy atomic
nucleus can be considered effectively immobile. Of course, such a
description is vitiated as soon as the momentum transfers become of the
order of the heavy mass, but then an expansion in powers of the inverse
mass works well~\cite{isgur}.

Until recent years, heavy quark fragmentation functions have been outside
the scope of experimental verification. Theoretical predictions for
fragmentation into light flavor baryons have been based on a range of
phenomenological models~\cite{peterson,hoodbhoy} and 
Monte-Carlo simulations~\cite{lund}. Subsequently, spin dependent
fragmentation functions were put on a firm foundation employing light-cone
field theory~\cite{ji} and the theoretical implications of the heavy quark
effective theory were incorporated~\cite{randall}. As these developments
occurred it was realized that the masses of the heavy flavor quarks allow
for perturbative calculations in the case of the doubly heavy
mesons~\cite{chang,braaten}. Later we extended these calculations to
include the heavy quark fragmentation of 
baryons~\cite{adamov1,adamov2}. (At the same time a Russian
group~\cite{russians,russians2} made a similar extension to baryons. More
recently the light cone expansion of the fragmentation functions was used
by the Amsterdam group~\cite{mulders} with an algebraic model to
generate predictions similar to ours.) The results of our phenomenological
approach were interesting for several reasons.
Working in the rather general parameter space we obtained the result that
the direct production rate of the spin excited baryons is on the order of or
higher than that of the ground state baryons of the same flavor~\cite
{adamov2}. We also found a functional dependence quite distinct from the
simple Peterson parameterization~\cite{peterson}.

That excited baryon result may turn out to be a sensible way to explain
the tendency seen throughout the scarce experimental data to indicate the
high depolarization of the heavy flavor baryons. If most of the singly heavy
baryons are produced in the ground state, then fragmentation can be seen as
the heavy quark picking up a couple of light sea quarks. Since the momentum
transfers associated with the quark fragmentation are not large enough to
make the heavy quark flip its spin, the expected depolarization is minimal,
i.e. the spin of the heavy baryon is expected to be nearly the same as that
of the heavy quark. However, if we include spin excited baryons into
consideration, the resulting polarization of the final baryon is less clear.
The excited states will decay strongly into the ground state almost entirely
via pion emission (with the exception of one radiative decay for the
$\Xi_c(\frac{1}{2}^\prime)$) and mix with the directly fragmented
sample.

It is our purpose herein to explore the fragmentation into excited heavy
baryons which then decay and populate the observable ground state
fragmentation. Since we are interested in the spin dependent fragmentation
functions for ground state heavy baryons, the intermediate state resonance
production will have a significant effect. In the next section we present
the particulars of the model calculation, incorporating QCD, heavy quark
approximations and the quark-diquark model of the baryons. The spin
dependent functions are developed and discussed in the following section,
including the relevant baryon wave functions in the quark-diquark model,
along with the comparison with data, where available. The final section
discusses the implications of these results in a broader context. An
Appendix presents the many scalar products that arise in the final
sum over states.

\section{Calculation details.}

\begin{figure}
\vspace{0.5in}
\includegraphics{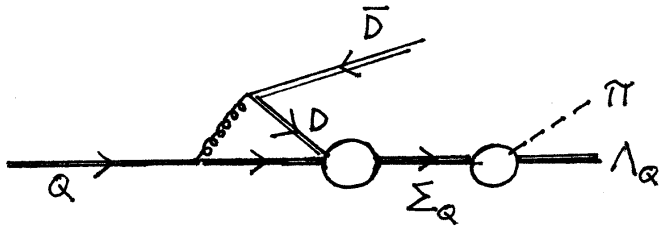}
\caption{Diagram representing the combined perturbative and soft QCD
calculation of fragmentation through excited baryons.}
\label{fig:diagram}
\end{figure}

The indirect fragmentation process is shown in the
Fig.~\ref{fig:diagram}.
The heavy quark
emits a hard gluon that splits into the diquark pair. Diquarks can be either
scalar or vector, with the vector diquark having the larger mass (as
suggested by the nucleon-$\Delta $ mass difference). If the diquark is
scalar, the heavy quark forms a ground state spin $\frac 12$ baryon with it.
Since the scalar diquark is spinless and the ground state has no orbital
angular momentum, the heavy baryon has nearly the same
helicity as the heavy quark. Vector diquarks have spin 1 and the resulting
helicity of the baryon is not the same as that of the single heavy quark,
but naturally depends on the helicity of the diquark. That does not produce
any depolarization of the heavy quark by itself - the heavy quark inside of
the baryon does not change its helicity. However, depolarization may occur
during the ``relaxation'' of the baryon into the ground state. The
depolarizing features of the spin excited baryon decay into the ground state
were discussed by Falk and Peskin \cite{peskin}. Two parameters are crucial
in determining the amount of depolarization relaxation. Let $\Delta M$ be
the mass splitting between the $\frac 32$ and $\frac 12^{\prime}$ excited
states of the heavy baryon. So $\Delta M$ will be related to the
spin-dependent part of the QCD interaction (which has a scale determined by $%
1/m_Q$) that, in turn, is responsible for flipping the quark spin. The
excited baryon states lifetimes will be nearly the same, $\frac 1\Gamma $.
The longer this lifetime is, in comparison to the time required for heavy
quark spin flipping, the more likely it is for the diquark and quark to mix
together forming a randomized quantum spin state. The time it would take for
the heavy quark to flip its helicity is proportional to $\frac 1{\Delta M}$.
The relative magnitude of these two parameters determines the nature of the
decay. Three distinct regions in the parameter space can be identified:

1) $\Gamma >>\Delta M$. In this case the heavy quark would not have enough
time to ``mix'' with the diquark or light degrees of freedom and the ground
state baryon helicity would be the same as the initial helicity of the heavy
quark.

2) $\Gamma <<\Delta M$. The decay proceeds very slowly, so the resulting
ground state baryon is completely depolarized..

3)$\Gamma \approx \Delta M$. This is the case of partial decoherence, the
borderline between cases 1 and 2.

It is important to note that the decay width $\Gamma $ is independent of the
heavy quark mass. Indeed, in a simplified model the decay may be due to the
diquark and light degrees of freedom only. On the other hand the mass
splitting $\Delta M\propto \frac{\Lambda _{QCD}^2}{m_Q}$ is inversely
proportional to the mass of the heavy quark. This shows that in the limit of
an infinitely heavy quark mass there is no depolarization. However, for any
given finite mass of the heavy quark all of the above possibilities may
occur.

There is very little experimental data on the absolute magnitude of the
decay widths for the heavy flavor excited baryons, but theoretical
predictions based on the heavy quark effective theory and potential models
(see Ref.~\cite{kwong} and the summary of Falk and
Peskin~\cite{peskin}) indicate that the decay
width is of the same order of magnitude as the mass splitting for
$\Sigma_b$ and smaller than the mass splitting for $\Sigma_c$.
That supports baryon relaxation as the source for the spin
depolarization. We will return to this issue after presenting an outline
of the calculation. 

We now proceed with the perturbative calculation of the fragmentation
functions. For simplicity we will first consider the spin independent
fragmentation function $f_1(z)$ (we use the notation of ref.~\cite{ji} but
without the carat in $\hat{f}_1$) . Once the formalism is established and
the calculations are made, it will be easier to turn to the case of the
spin dependent $h_1$ and $g_1$ in the next section. The partial
width~\cite{braaten} for the inclusive decay process $Z^0\rightarrow H+X$
can be written in general for any hadron $H$ as 
\begin{equation}
d\Gamma (Z^0\rightarrow H(E)+X)=\sum_i\!\int_0^1\!dz\,d\hat{\Gamma}%
\!(Z^0\rightarrow i(E/z)+X,\mu )\,D_{i\rightarrow H}\!(z,\mu ),
\label{eq:general}
\end{equation}
where $H$ is the hadron of energy $E$ and longitudinal momentum fraction
$z$ relative to the parton {\it i}, while $\mu $ is the arbitrary
renormalization scale whose value will be chosen to avoid large logarithms.
The function $D_{i\rightarrow H}$ is the fragmentation probability in a
paricular channel. The parton $i$ is a heavy quark and the baryon $H = B$
could be $\Lambda_Q, \Sigma_Q,$ or $\Xi _Q$ in our model. In order to
extract the leading twist fragmentation function $f_1(z,\mu )$ out of the
total fragmentation probability we take the high energy limit. More
precisly the limit is of the proper light cone component $l_0+l_3$. As
long as we are only looking at leading twist functions, the high
energy limit produces the same result.
The expression, Eqn.~\ref{eq:general} will be simplified if we restrict
ourselves to the fragmentation channel shown in
Fig.~\ref{fig:diagram}. Then, while $z$ is kept fixed, the limit of
large mass of the $Z^0$ along with large energy of the heavy quark, $q_0$,
and the baryon, $l_0$, yields 
\begin{equation}
\lim_{l_0\longrightarrow \infty }d\Gamma (Z^0\rightarrow
B(E)+X)=\lim_{q_0\longrightarrow \infty }\int_0^1\!dz\,d\hat{\Gamma}%
\!(Z^0\rightarrow Q(E/z)+X,\mu )\,f_1(z,\mu ).  \label{simple}
\end{equation}
and
\begin{equation}
\int_0^1\!dz\,f_1(z,\mu )=\frac{\Gamma _1}{\Gamma _0},  \label{f1}
\end{equation}
where $\Gamma _1$ is the decay width of $Z^0$ into the ground state baryon
and appropriate remnants - antiquark, spectator diquark and pion, while
$\Gamma_0$ is the total decay width of the $Z^0$ into the heavy quark
pair~\cite{adamov1}.

The decay width of the (infinitely massive) $Z^0$ into the inclusive heavy
ground state baryon $B_Q$ is viewed as a direct decay into $Q + \bar{Q}$
wherein the quark is off-shell and subsequently fragments. The
fragmentation is finally into the $B_Q$, an anti-diquark $\overline{D}$
and a single pion (or photon) from the intermediate resonant baryon
decay and is represented by the integral: 
\begin{equation}
\Gamma _1=\frac 1{2M_Z}\int [d\overline{q}][dl][dp^{\prime}][d\overline{\pi 
}](2\pi )^4\delta ^4(Z-\overline{q}-l-p^{\prime }-\overline{\pi })\left|
M_1\right| ^2  
\label{gam1}
\end{equation}
where $\bar{q}$, $l$, $p^{\prime }$ and $\pi $ are the 4-momenta of the $%
\bar{Q}$, $B_Q$,$\overline{D}$ and the pion (or photon), respectively, and
the amplitude 
$M_1$ is summed and averaged over unobserved spins and colors. We use the
notation $[dp]=d^3p/(16\pi ^3p_0)$ for the invariant phase space element. To
isolate the fragmentation function, the production of the fragmenting quark (%
$d\hat{\Gamma}$ of Eq.~\ref{eq:general}) must be factored out. The
fictitious decay width for the $Z^0\rightarrow Q+\bar{Q}$, with the 
$Q$-quark on shell is 
\begin{equation}
\Gamma _0\,=\,{\frac 1{2M_Z}}\!\int \![d\bar{q}][dq]\,(2\pi )^4\delta ^4(Z-%
\bar{q}-q){\frac 13}\!\sum \!|M_0|^2.  \label{eq:gam0}
\end{equation}
with $q$ the heavy quark 4-momentum.

In order to factor the fictitious decay width out of Eq.~\ref{gam1} we have
to transform the phase space variables. That can be achieved by
introducing new,
production independent variables $x_1=\frac{p_0+p_L}{q_0+q_L}$ and 
$x_2=\frac{l_0+l_L}{p_0+p_L}$ that can be loosely thought of as Feynman scaling
variables for each subprocess, i.e. excited baryon production and decay. We
will introduce a further simplification - the ratio of the narrow decay
width to the mass of the excited baryons allows the narrow width
approximation to be used. The square of the denominator for each excited
state propagator that enters the squared amplitude can be factored out for
each decay channel and is proportional to $\frac 1{(p^2-M^2)+M^2\Gamma^2}$.
Hence, in the limit of the small decay width to mass ratio, that
factor can be approximated by $\frac \pi {M\Gamma }\delta (p^2-M^2)$,
effectively putting the excited baryon back on the mass shell. The cross
terms that come from the cross products of the propagators of the 
${\frac 12}^{\prime}$ and ${\frac 32}$ spin channels of the decay
disappear if the mass difference is much larger than the decay width.

The resulting phase space integral can be written as: 
\begin{eqnarray}
\Gamma _1 & = & \frac 1{2M_Z}\frac 1{256\pi ^4}\int [d\overline{q}][dq](2\pi
)^4\delta ^4(Z-q-\overline{q})  \nonumber \\
& & \cdot \int ds_q\theta \left( s_q-\frac{M_\Sigma ^2}z-\frac{m_d^2}{1-z}%
\right) \int d\phi d\varphi dx_1dx_2\left| A_1\right| ^2
\end{eqnarray}
Here $\left| A_1\right| ^2\delta (p^2-M^2)=\left| M_1\right| ^2$. The two
angles $\phi$ and $\varphi$ introduced here deserve some special
attention and are defined carefuly in the Appendix. They are associated
with the position of the transverse momentum
vector in two frames of reference. The first is the frame determined by the
three-momentum of the heavy quark and a fixed vector perpendicular to it,
which is arbitrary unless it is the quark's transverse spin vector (which
enters in $h_1$ only). The angle $\phi$ is the azimuthal angle between this
plane and the transverse momentum vector (relative to the heavy quark
direction) of the excited baryon. The second plane (or frame if we add the
vector perpendicular to the first two) is constructed out of the
three-momentum of the excited baryon and the spin vector perpendicular to
that three-momentum but having no transverse component relative to the first
frame. The second angle $\varphi$ is defined as the azimuthal angle between
this latter plane and the transverse momentum (relative to the excited
baryon direction) of the final baryon. 

The spin averaged matrix element $\left| A_1\right|^2$ has no angular  
dependences, so we can safely integrate
over the angles. That will not be true for the $g_1$ and especially $h_1$.
The choice of $x_1$ and $x_2$ helps to keep the integral symmetric looking,
but unlike the standard scaling variable $z=\frac{l_0+l_L}{q_0+q_L}$ these
variables are not experimentally observable. Using $z=x_1x_2$ we can
finally rewrite the phase space integral to be: 
\begin{eqnarray}
\Gamma _1 & = & \frac 1{2M_Z}\frac 1{16\pi^2}\int[d\overline{q}][dq](2\pi
)^4\delta ^4(Z-q-\overline{q})  \nonumber \\
& & \cdot \int ds_q\theta (s_q-\frac{M_\Sigma ^2}z-\frac{m_d^2}{1-z})\int 
\frac{dx_2}{x_2}dz\left| A_1\right| ^2
\end{eqnarray}

After factoring out the production decay width we are left with the somewhat
simpler expression for $f_1$:
\begin{equation}
f_1(z,\mu)={\frac 1{16\pi ^2}}\lim_{q_0\rightarrow \infty
}\int_{s_{th}}^\infty \!ds\frac{dx_2}{x_2}{\frac{\left| A_1\right|^2}{%
\left| M_0\right|^2}}  
\label{eq:ratio}
\end{equation}

The expression above is general for the four body final state. The model
dependence is hidden inside of the $\left| A_1\right|^2$ with the delta
function obtained in the narrow width approximation integrated out. The
final ground state baryon can be produced via one of the two intermediate
states. If the states are separated by a mass gap wider than the decay
width
they do no not interfere with each other. That is the scenario supported by
the theoretical predictions and the experimental data we have at the moment.
It follows that the two channels for the indirect baryon production can be
considered independently. The amplitudes for both of them can be
expressed as: 
\begin{equation}
A=\overline{U}(l)\left\{\frac{K_{1/2}(\not{p}+M_\Sigma
)A_{1/2}}{\sqrt{\Gamma
_{1/2}}}+\frac{K_{3/2}^\mu P_{\mu \nu }(\not{p}+M_\Sigma )A_{3/2}^\nu }{%
\sqrt{\Gamma _{3/2}}}\right\}\Pi   \label{eq:total}
\end{equation}
where $\Pi$ is the quark production spinor, $K$ is the decay operator
for $1/2^{\prime}$ (into a pion or a photon) or $3/2$ baryon, the
subscripted $A$ is the corresponding production
operator for the excited baryon and $P_{\mu\nu}(\not{p}+M_\Sigma)$ is
the spin sum of the 3/2 baryon.
\begin{eqnarray}
K_{1/2} &=&g_1(\not{p}-\not{l})\gamma _5 \nonumber \\
K_{3/2}^\nu  &=&-g_2(p-l)^\nu \nonumber \\
P_{\mu \nu } &=&-g_{\mu \nu }+\frac 13\gamma _\mu \gamma _\nu +\frac
1{3M}(\gamma _\mu p_\nu -\gamma _\nu p_\mu )+\frac 2{3M^2}p_\mu p_\nu
\label{eq:decays}
\end{eqnarray}
where $g_1$ and $g_2$ are coupling constants associated with the
decays. Their exact value is irrelevant in this narrow width approximation
for the resonance that decays into a single channel -- both denominator
and numerator of Eq.\ref{eq:total}
contain the second power of the decay coupling constant.

Note that $\Xi_c$ has to be treated differently. The spin excited
states are widely separated, so the lowest lying spin ${\frac 12}^\prime
\Xi_c(2574)$ state can no
longer decay via $\pi$. That only leaves the possibility of photon
decay with a branching ratio of nearly 100\%. For the radiative decay we
will consider the  simple electric dipole transition amplitude,
$g_3\overline{U}(l)\not{\!\epsilon}(\gamma)U(p)$, where the photon's
polarization vector $\epsilon(\gamma)$ enters. This leads to a $K_{1/2} =
g_3\!\not{\!\epsilon}$ replacing the value in Eq.~\ref{eq:decays}. 

The ground state baryon is composed of a heavy quark and a scalar
diquark. For direct production of the ground state there is
one coupling constant for the scalar diquark to couple to the gluon field
via a color octet vector current -- a color charge
strength, along with a possible form factor $F_s$. 
\begin{equation}
J_{\mu}^{A(S)} = g_s F_s(k^2) (p+p')_{\mu}S^{\alpha
\dagger}\lambda_{\alpha \beta}^{A}S^{\beta},
\label{eq:scalar}
\end{equation}
where $p$ and $p'$ are the scalar diquark 4-momenta and $k=p'-p$. Then the
amplitude for direct production becomes
\begin{equation}
A_{S\,1/2}=-\frac{\psi(0)}{\sqrt{2m_d}}F_S(k^2)\bar{U}_Bg_s
[k_{\lambda}-
2m_d v_{\lambda}]P^{\lambda},
\label{eq:As}
\end{equation}
where
\begin{equation}
P^{\lambda}=
\bigtriangleup^{\lambda \nu}g_s\gamma_{\nu}
\frac{m_Q(1+\mathbf{v})+\mathbf{k}}{(s-m^2_Q)}\Gamma.
\label{eq:Plambda}
\end{equation}

The amplitudes for the production of the ${\frac 12}'$ and ${\frac 32}$
states involve the gluon coupling to the vector diquark. For the vector
diquark, the color octet current (which couples to the gluon field
vector) is more complicated~\cite{gold1}. There
are three constants - color charge, anomalous chromomagnetic dipole
moment $\kappa$, and chromoelectric quadrupole moment $\lambda$, along
with the corresponding form factors, $F_E,\,F_M,\ {\rm and}\ F_Q$,
\begin{equation}
\begin{array}{rcl}
J_{\mu}^{A(V)} & = & g_s(\lambda^A)_{\beta
\alpha}\left\{F_E(k^2)[\epsilon^{\alpha}(p)
\cdot\epsilon^{\beta \dagger}(p')](p+p')_{\mu}\right.\\
   & & +(1+\kappa)F_M(k^2)[\epsilon_{\mu}^{\alpha}(p)p
\cdot\epsilon^{\beta \dagger}(p')+\epsilon_{\mu}^{\beta
\dagger}(p')p'\cdot
\epsilon^{\alpha}(p)]\\
   & &
 +\frac{\lambda}{m_D^2}F_Q(k^2)[\epsilon_{\rho}^{\alpha}(p)
\epsilon_{\nu}^{\beta \dagger}(p')+\frac{1}{2}g_{\rho \nu}
\epsilon^{\alpha}(p)\cdot\epsilon^{\beta
\dagger}(p')]k^{\rho}k^{\nu}(p+p')_{\mu}\left.\right\},
\end{array}
\label{eq:vector}
\end{equation}
where $A$ is the color octet index, $\alpha,\,\beta$, ..., are color
anti-triplet indices, the $\epsilon$'s are polarization 4-vectors for
the diquarks. The chromoelectric part of the matrix element contributing
to the spin ${\frac {1}{2}}^\prime$ baryon is
\begin{equation}
A_{E\,1/2}=-\frac{\psi(0)}{\sqrt{3m_d}}F_E(k^2)
\gamma_5\gamma^{\mu}
\frac{1+\mathbf{v}}{2}g_s\bar{\epsilon}^{*}_{\mu}[k_{\lambda}-
2m_d v_{\lambda}]P^{\lambda}.
\label{eq:E1/2}
\end{equation}
The chromomagnetic contribution to the spin ${\frac {1}{2}}^\prime$
baryon is taken to be
small based on earlier extimates from baryon
spectroscopy~\cite{adamov1,adamov2,gold1}. The quadrupole coupling is
assumed inconsequential due to the more rapid fall-off with momentum
transfer from dimensional counting rules. For the spin $\frac{3}{2}$ 
baryon the corresponding chromoelectric amplitude is 
\begin{equation}
A_{E\,3/2}^{\nu}=-\frac{\psi(0)}{\sqrt{2m_d}}F_E(k^2)
g_s\bar{\epsilon}^{*\nu}[k_{\lambda}-
2m_d v_{\lambda}]P^{\lambda},
\label{eq:E3/2}
\end{equation}

After some simplification we can write the above equations including the
decays:
\begin{eqnarray}
\overline{U}(l)K_{1/2}(\not{p}+M_\Sigma )A_{1/2}&=&-\frac{\psi
(0)(M_\Sigma
+M_\Lambda )g_{1/2}}{\sqrt{6m_d}}F_E(k^2)\frac{2g_s^2}{M_\Sigma
(s-m_Q^2)} \nonumber \\
  & &\overline{U}(l)(\not{p}-M_\Sigma ) 
(M_\Sigma \not{\epsilon}^{*}+(\epsilon ^{*}p)) \nonumber \\
  & &[2M_\Sigma^2(1-r)-2\frac{(np)}{(nk)}(kp)+M_\Sigma \not{k}]
\end{eqnarray}
for $\pi$ decay of the 1/2$^\prime$;
\begin{eqnarray}
\overline{U}(l)K_{1/2}(\not{p}+M_\Sigma )A_{1/2}&=&-\frac{\psi
(0)(M_\Sigma
+M_\Lambda )g_{1/2}}{\sqrt{6m_d}}F_E(k^2)\frac{2g_s^2}{M_\Sigma
(s-m_Q^2)} \nonumber \\
  & &\overline{U}(l)\gamma_5\not{\epsilon}(\pi^2)(\not{p}-M_\Sigma )
(M_\Sigma \not{\epsilon}^{*}+(\epsilon ^{*}p)) \nonumber \\
  & &[2M_\Sigma^2(1-r)-2\frac{(np)}{(nk)}(kp)+M_\Sigma \not{k}]
\end{eqnarray}
for the photon ($\epsilon(\pi^2)$) decay of the 1/2$^\prime$;
\begin{eqnarray}
\overline{U}(l)K_{3/2}^\mu P_{\mu \nu }(\not{p}+M_\Sigma)A_{3/2}^\nu &=&
{\frac{\psi (0) M_\Sigma g_{3/2}}{{\sqrt{2m_d}}}F_E(k^2)
\frac{2g_s^2}{M_\Sigma(s-m_Q^2)}}
\nonumber \\
    & &{\overline{U}}(l)(p^\mu -l^\mu )P_{\mu\nu}(\not{p}+M_\Sigma)\epsilon^{*\mu}
\nonumber \\
  & &  \lbrack 2M_\Sigma ^2(1-r)-2\frac{(np)}{(nk)}(kp)+M_\Sigma \not{k}]
\end{eqnarray}
for the $\pi$ decay of the 3/2 state,
where $r = m_d/M_{\Sigma}$ is a measure of the departure from the heavy
quark limit. Note that the $\Xi_c(\frac{1}{2}^\prime)$ radiative decay
is somewhat different from $K_{1/2}$ above, but easily accommodated in
the sum over intermediate states.

The fragmentation functions must be
independent of the heavy quark production mechanism. For
simplicity then, we actually calculate the decay of a heavy scalar
breaking into the heavy quark antiquark pair with the quark further
fragmenting into the baryon. The remaining calculations are
straightforward, but quite complex. We will sketch them in the next
section. The results for the different heavy quark flavors are presented
following that.

\section{Spin dependent fragmentation functions.}

The leading twist fragmentation functions $g_1$ and $h_1$ carry valuable
information about the helicity and transversity transfer in the
system. The meaning of these functions first introduced by Jaffe and
Ji~\cite{ji} is straightforward. The function $g_1$ is the difference
between probabilities for having the
final baryon with  helicity aligned with and opposite to the original
helicity of the heavy quark; $h_1$ similarly represents the difference in
the no flip and flip rates, but in the sense of the transverse direction
of spin (more carefully, the transversity~\cite{transversity}).

In the strict limit of the heavy quark effective theory one expects to find 
$f_1(z)=g_1(z)=h_1(z)=P\delta (1-z)$, with $P$ being an overall production
rate for the corresponding baryon. This is not quite so for any given finite
mass of the heavy quark. The only restrictions come from the probabilistic
interpretation of the fragmentation function (for example, 
$f_1(z)\geq  g_1(z)$) and the analog of the proposed structure function 
inequality~\cite{soffer}
$f_1(z)+g_1(z)\geq \left|2h_1(z)\right|$. Both of these restrictions are  
satisfied in our model.

The theoretical prediction made in ref.~\cite{peskin} indicates that the
polarization of the $\Lambda_Q$ is heavily dependent on the production
rates for the excited baryons. We only consider the lowest spin excited
baryons $\Sigma_Q(\frac{1}{2}^\prime)$ and $\Sigma_Q^{*}(\frac{3}{2})$ in
the present paper. Other  
excitations can be included in two ways: considering excited
diquarks, either radially or orbitally, or considering  quark-diquark
baryon configurations excited radially and/or orbitally. The
former case creates a lot of theoretical uncertainties, since the diquark
is less bound and can hardly be considered as a parton. In the latter
case, the wave functions at the origin
are expected to be smaller, so that the production rate of such states is
smaller. In general, such states may provide for corrections to our
calculations, but the main features of the fragmentation should remain
unaltered.

In our model the spin dependent fragmentation functions can be obtained
directly, using the modified Eq.\ref{eq:ratio}: 
\begin{eqnarray}
g_1(Q,z) &=&{\frac 1{256\pi ^4}}\lim_{q_0\rightarrow \infty
}\int_{s_{th}}^\infty \!ds\frac{dx_2}{x_2}d\phi d\varphi {\frac{\left|
A_{1+}\right| ^2-\left| A_{1-}\right| ^2}{\left| M_0\right| ^2}}
\label{eq:spin} \\
h_1(Q,z) &=&{\frac 1{256\pi ^4}}\lim_{q_0\rightarrow \infty
}\int_{s_{th}}^\infty \!ds\frac{dx_2}{x_2}d\phi d\varphi {\frac{\left|
A_{1y+}\right| ^2-\left| A_{1y-}\right| ^2}{\left| M_0\right| ^2}}
\label{eq:spinh1} 
\end{eqnarray}
with new indices specifying the spin alignment ($|y+> \sim |+>+i|->$).
Angular integration is
especially complicated for $h_1$, because matrix elements involve spin
projections that make them no longer azimuthally symmetric. Naturally, the
transverse spin vector would set a preferred azimuthal direction that
manifests itself in the scalar products of the transverse spin vectors 
with other vectors that have transverse components. These scalar products
are listed in the Appendix.

The amplitude $A$ is defined in Eq.~\ref{eq:total} and in general can be
represented as:
\begin{equation}
A=\overline{U}O\Pi 
\end{equation}
where $O$ is the expression in curly brackets in Eq.~\ref{eq:total}.
The square of the matrix element can then be written as:
\begin{equation}
\left| A_{1\alpha }\right| ^2=Tr(\Pi \overline{\Pi }\gamma _0O^{\dagger
}\gamma _0(\not{p}+M_\Lambda )\frac{1+\gamma _5\not{S}_\alpha }2O)
\label{eq:trace}
\end{equation}
with $\alpha $ being the spin projection index and $S_\alpha $ being the
spin four-vector corresponding to that projection. The two cases for
this problem are the longitudinal or helicity and transverse spin
vectors. The longitudinal spin vectors are defined unambiguously by
defining the four-vector in the rest frame of the particle with time
component equal to zero and space component equal to the unit vector
pointed in the direction of the Lorentz boost that would take the particle
back into the lab frame. The transverse spin vector
(transversity~\cite{transversity}) has to
be perpendicular to the direction of the Lorentz boost and has zero energy
component. It will not be affected by the
Lorentz boost, so it will still be perpendicular to the three-momentum of
the particle even in the lab frame. The tip of the spin vector could be
anywhere on the unit circle that has its origin at the base of the spin
vector and is perpendicular to the three-momentum of the particle. The
ambiguity is resolved by choosing the positive direction of the transverse
spin of the final baryon  to ``align'' (or anti-align) with the
arbitrarily chosen positive direction of the initial quark's
transversity. By taking the three-momentum of the quark to
be along the z-axis and transverse spin direction to be along the x-axis
the transverse momentum will be defined to be in the xz plane with it's x
component positive. Such a definition of the transverse spin vector also
naturally introduces the new coordinate system with z along the three
-momentum of the baryon and the x-axis parallel to the transverse
spin.  In exactly the same fashion we can introduce the transverse spin
vector of the intermediate baryon. We set up yet another
coordinate frame. The angles $\phi $ and $\varphi $ are azimuthal angles
of the intermediate baryon in the frame of the heavy quark and the final
baryon in the frame of intermediate baryon (see the Appendix). 

As was mentioned previously, we can proceed in
explicitly calculating all matrix elements without losing any generality by
assuming that the production mechanism is a simple decay of the scalar
particle. This way $\Pi =\upsilon _{-\alpha }$, where $-\alpha $ indicates
that the spin of the antiquark is opposite of the expected spin of the
quark. After simplifying the squares of matrix elements and leaving only
the leading terms in the high momentum limit, we end up with the
scalar products of all the involved four-momenta and spin vectors that 
are given in the Appendix. We next
look at the angular composition of the resulting integrands.

In general, the square matrix elements involve scalar products of the
available four-vectors:

$\overline{q},q,p,l,n,s^q,s^{\overline{q}},s^l$.

The spin vectors can be either transverse or
longitudinal and $n$ is the four-vector orthogonal to the quark momentum
that enters in the axial gauge. The phase space integration variables are 

$x_1, x_2, \phi ,\varphi ,s_q$.

We will use the notation $p_{qt}$ throughout this paper, where the first
index stands for the three-momentum that originated the frame and
the second index, if there, corresponds to the component of the vector. So
$p_{qt}$ is the transverse component of the excited baryon three-momentum
viewed in the first frame. Note that 
\begin{equation}
p_{qt}^2=s_qx_1(1-x_1)-M_\Sigma ^2(1-x_1)-m_d^2x_1,
\end{equation}
\begin{equation}
l_{pt}^2=M_\Sigma ^2x_2(1-x_2)-M_\Lambda ^2(1-x_2)-m_d^2x_2.
\end{equation}

In the Appendix we consider each scalar product that can arise in the
integrands of Eq.~\ref{eq:spin} and~\ref{eq:spinh1}. 

For the $f_1$ there are no spin vectors since we spin average everything.
This way the only scalar product that will introduce angular dependence
is $(ql)$ (as shown in the Appendix), the product of the four-momenta of
the initial quark and
baryon. The angular dependence is proportional to $\cos (\varphi +\phi
)$. This term disappears after the single integral over one of the angles
-- introducing intermediate baryon degrees of freedom breaks the azimuthal
symmetry of the system, but the integration restores the symmetry.

The function $g_1$ has the same type of angular dependence. The only
scalar products that produce azimuthal angles are $(ql)$, $(s_ql)$ and
$(s_lq)$. All of them are
proportional to $\cos (\varphi +\phi )$ and the angular dependence is
removed by integrating over one of the angles, which is not surprising given
that the azimuthal symmetry should be preserved again.

In the case of $h_1$ the situation becomes more complex. The scalar products
of the type $(s_qp)$ and $(s_lp)$ are proportional to $\cos (\phi )$ and 
$\cos (\varphi )$ and we still have the scalar product $(ql)$ that is
proportional to $\cos(\varphi +\phi )$. This type of angular dependence  
is removed only after both angles are integrated. The system is never
azimuthally symmetric -- in choosing the arbitrary transverse spin
direction we break the rotational invariance. 

In order to obtain the final fragmentation function we only have to
integrate over the angles and change all the variables into $x_1$ and
$x_2$, which can be done in the high momentum limit. This integration is
performed numerically. There are several parameters that need to be
specified. The masses of the diquarks are taken as (uu, ud) 0.6
GeV/c$^2$ and (us, ds) 0.9 GeV/c$^2$. The quark masses are 
4.9 GeV/c$^2$ for the b-quark and 1.6 GeV/c$^2$ for the c-quark.
The $\Lambda_b$ mass is 5.6 GeV/c$^2$; the $\Sigma_b$ mass is 5.8
GeV/c$^2$ (the mass of the $\Sigma_b*$ can be taken the same,
because difference is in the hundredths); the $\Lambda_c$ is 2.3
GeV/c$^2$; the mass of $\Sigma_c$ is 2.5 GeV/c$^2$. The wave functions for
the formation of the baryons are obtained from the power law potential of
Eichten and Quigg~\cite{quigg}, 
\begin{equation}
V(r)=-8.064\ {\rm GeV} + 6.898\ {\rm GeV}\ (r\times 1\ {\rm GeV})^{0.1}.
\end{equation}
The resulting square moduli of the wavefunctions at the origin,
$|\psi(0)|^2$, are 0.46 GeV$^3$, 0.35 GeV$^3$ and 0.51 GeV$^3$ for
$\Lambda_b, \Lambda_c$ and $\Xi_c$, respectively. 

The integrations produce spin-dependent fragmentation functions that are
defined at the
scale $\mu_0 = m_Q + m_{diquark}$. To evolve them to higher values of the
defining scale (or the typical $Q^2$) we utilize the appropriate
spin-dependent Altarelli-Parisi integro-differential equations as
determined by Artru and Mekhfi~\cite{artru}. 

\begin{figure}
\includegraphics{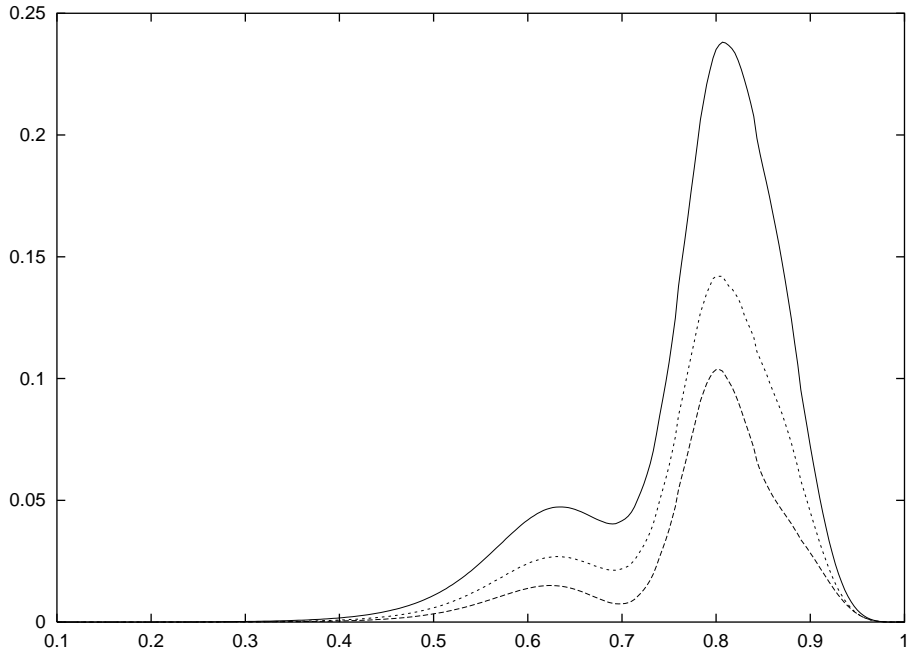}
\vspace{2.5in}
\caption{Fragmentation functions for $\Lambda_c$ at $\mu_0$. At the peak
$f_1$ is largest, followed by $h_1$ and $g_1$.}
\label{fig:c_no}
\end{figure}

\begin{figure}
\includegraphics{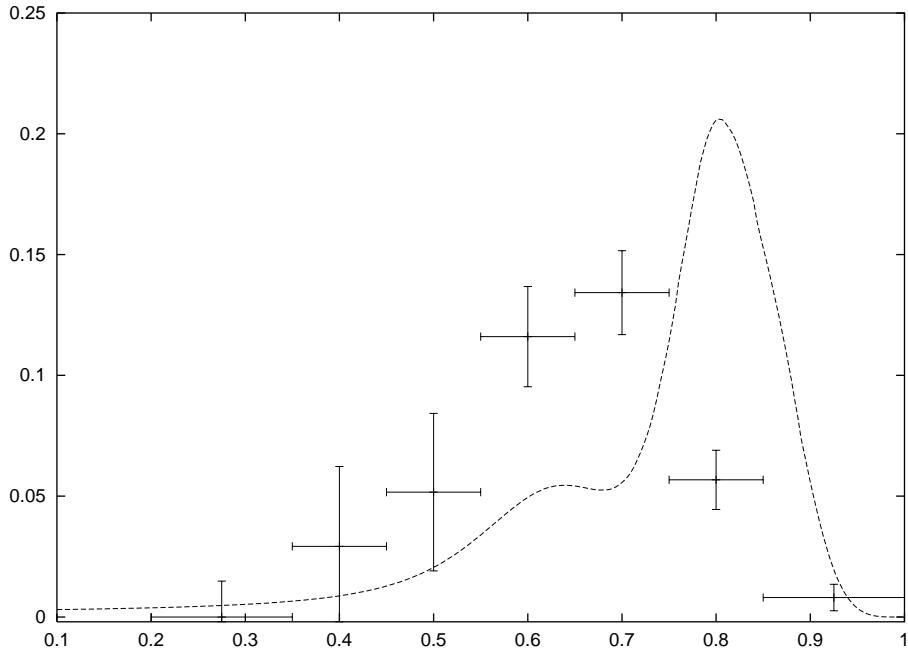}
\vspace{1.7in}
\caption{Spin independent function $f_1(z)$ evolved to $\mu$= 5.5 GeV. The
data are from CLEO~\cite{cleo1}.}
\label{fig:c_data}
\end{figure}

In Fig.~\ref{fig:c_no} the unevolved fragmentation functions $f_1, g_1,
h_1$ are plotted for $\Lambda_c$ as functions of $z$. The function
$f_1(z)$ evolved to 5.5 GeV (half of the CESR energy) is shown in
Fig.~\ref{fig:c_data}, along with the CLEO data~\cite{cleo1} on
$\Lambda_c$ production from one of the decay channels,
$pK^-\pi^+$. The data are given for bins of $x_+$, which is related to our
$z$, but somewhat different at finite c-quark energy. The normalization of
the data (which is arbitrary) was adjusted to correspond to the
normalization of the predicted curve. There is considerable scatter about
our curve for $f_1(z)$. A smooth curve through the data suggests a lower
$z$ for the peak position. The data are not sufficiently spaced in $x_+$
to check whether or not there is evidence for the shoulder in our
predicted curve. Caution is advised in interpreting the CLEO $\Lambda_c$ 
production data as a leading twist fragmentation function, given
that the c-quark is being produced at half the $\Upsilon$ mass. This is
far from a large energy compared to the fragmenting $\Lambda_c$,
i.e. $\frac{1}{2}M_{\Lambda_c}/M_{\Upsilon} \sim 0.4$ which is quite
sizeable. 

\begin{figure}
\includegraphics{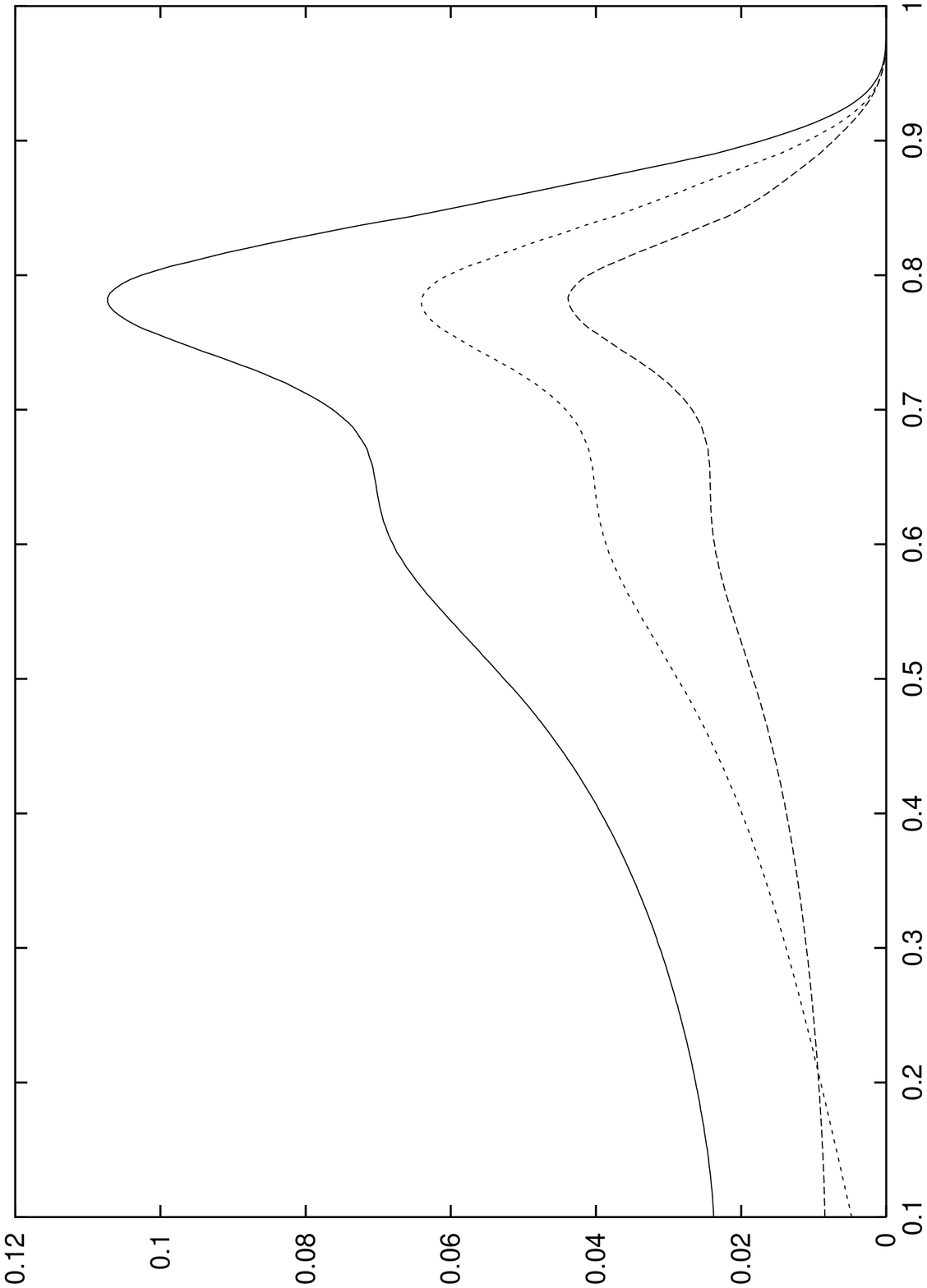}
\vspace{2.5in}
\caption{Fragmentation functions for $\Lambda_c$ evolved to $\mu$= 45
GeV. At the peak $f_1$ is largest, followed by $h_1$ and $g_1$.}
\label{fig:c_ev}
\end{figure}

\begin{figure}
\includegraphics{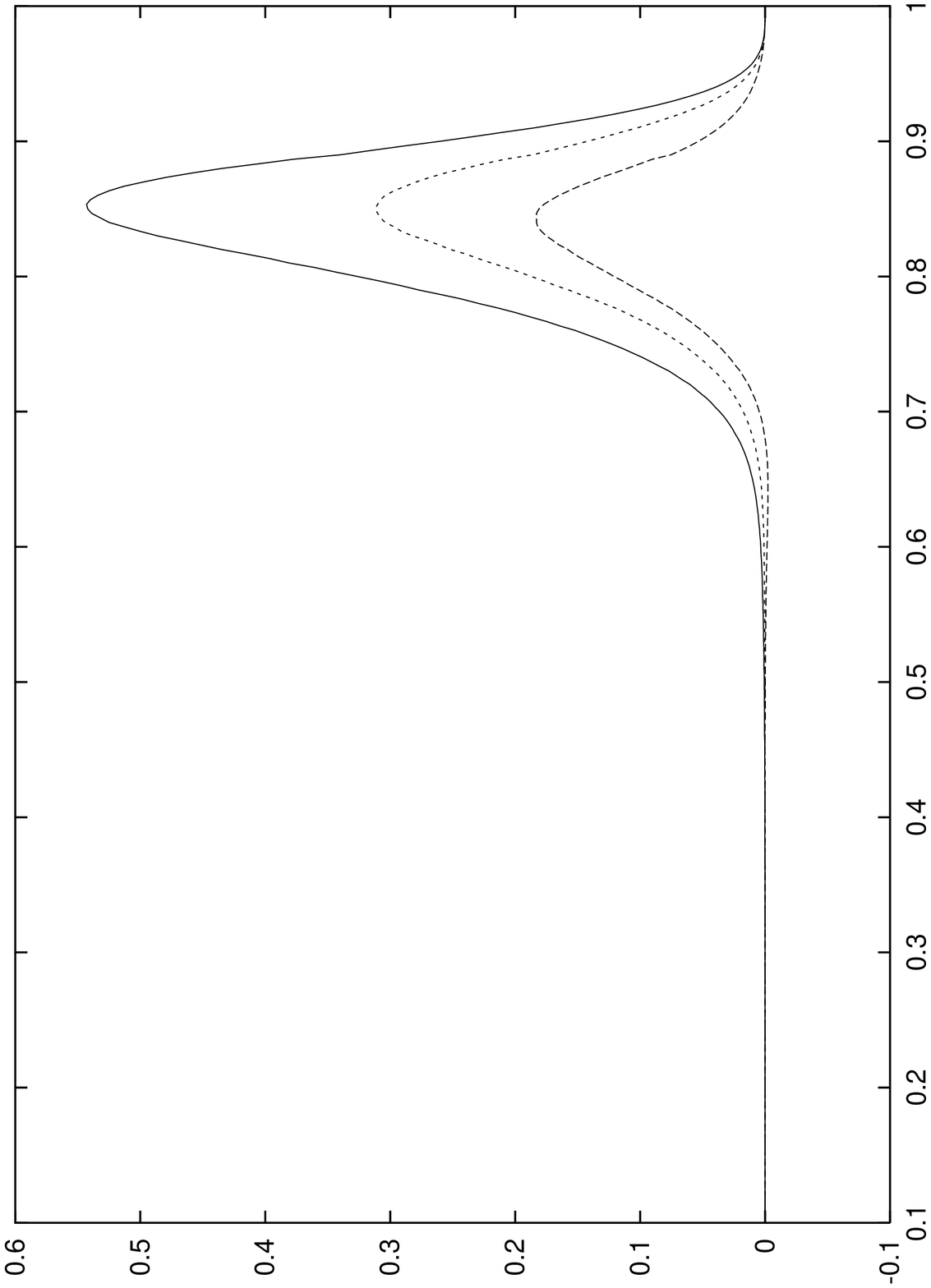}
\vspace{1.7in}
\caption{Fragmentation functions for $\Lambda_b$ at $\mu_0$. At the peak
$f_1$ is largest, followed by $h_1$ and $g_1$.}
\label{fig:b_no}
\end{figure}

\begin{figure}
\includegraphics{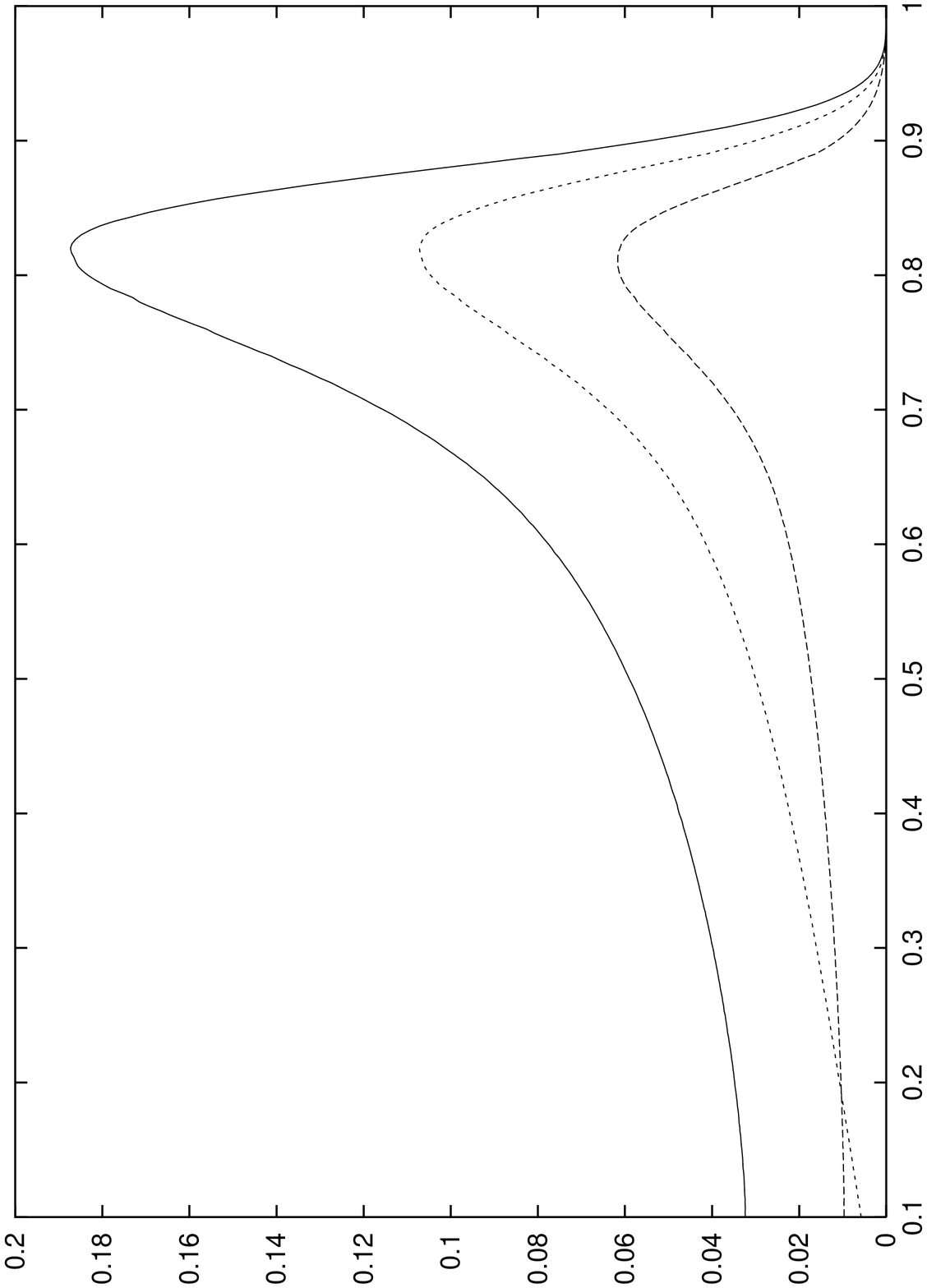}
\vspace{2.5in}
\caption{Fragmentation functions for $\Lambda_b$ evolved to 45 GeV. At the
peak $f_1$ is largest, followed by $h_1$ and $g_1$.}
\label{fig:b_ev}
\end{figure}

\begin{figure}
\includegraphics{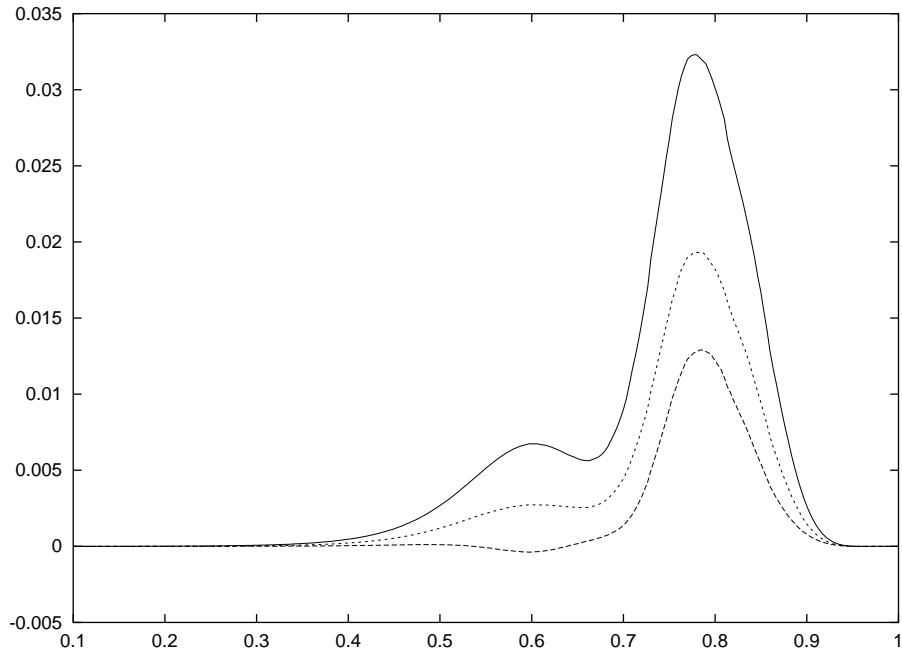}
\vspace{1.7in}
\caption{Fragmentation functions for $\Xi_c$ unevolved. At the
peak $f_1$ is largest, followed by $h_1$ and $g_1$.}
\label{fig:xi_c_no}
\end{figure}

\begin{figure}
\includegraphics{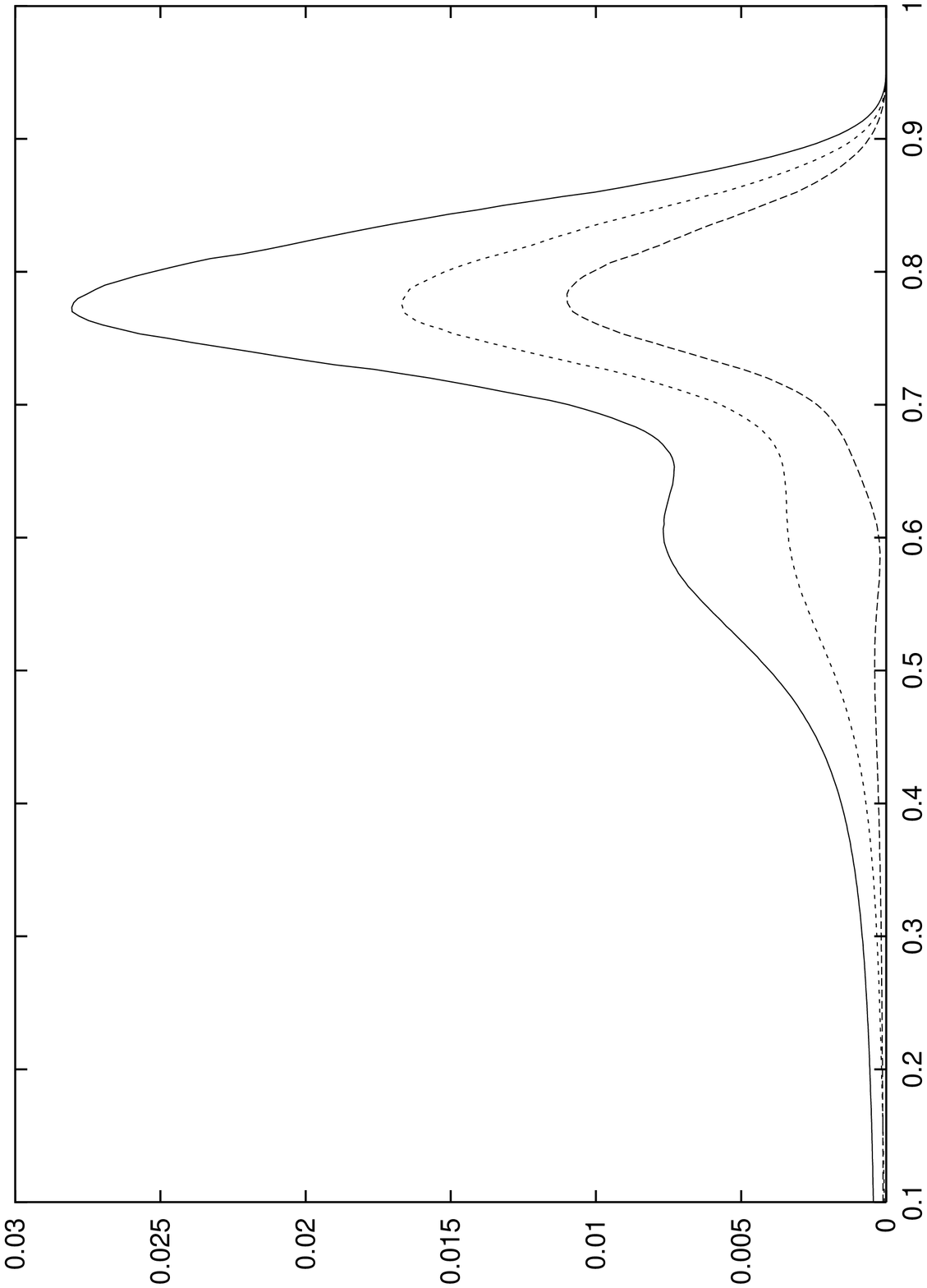}
\vspace{2.5in}
\caption{Fragmentation functions for $\Xi_c$ evolved to 5 GeV/c. At the
peak $f_1$ is largest, followed by $h_1$ and $g_1$.}
\label{fig:xi_c_5}
\end{figure}

\begin{figure}
\includegraphics{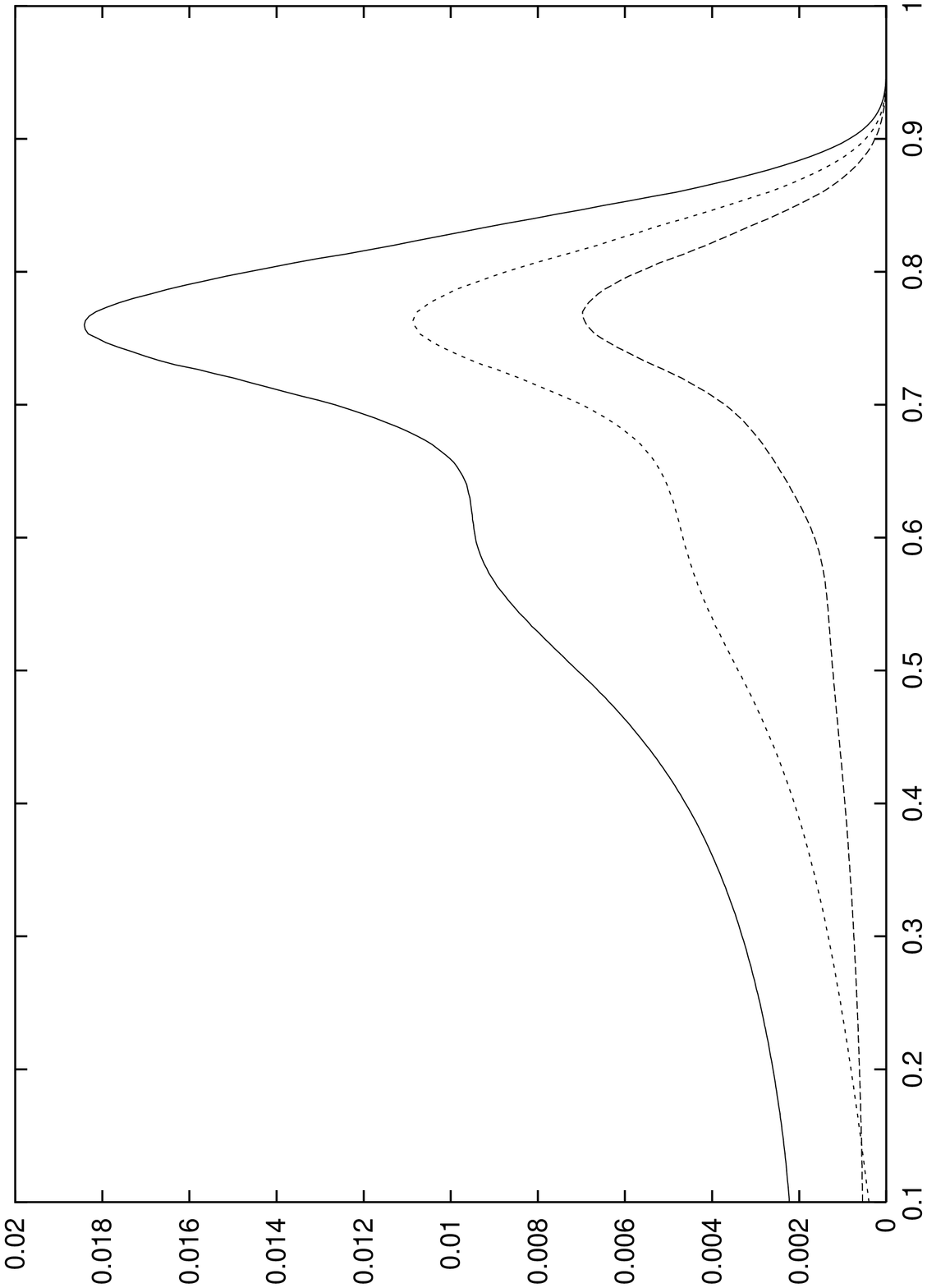}
\vspace{2.5in}
\caption{Fragmentation functions for $\Xi_c$ evolved to 45 GeV/c. At the
peak $f_1$ is largest, followed by $h_1$ and $g_1$.}
\label{fig:xi_c_45}
\end{figure}

The 45 GeV spin dependent functions are shown in Fig.~\ref{fig:c_ev} for
completeness. Corresponding results for the b-quark 
are shown for $\Lambda_b$ in Fig.~\ref{fig:b_no} and~\ref{fig:b_ev}. The
$\Xi_c$ fragmentation functions are presented in
Figs.~\ref{fig:xi_c_no},\ref{fig:xi_c_5} and \ref{fig:xi_c_45}, for the
unevolved, 5.5 GeV/c (for comparing with CESR data) and 45 GeV/c (for
comparing with LEP data).

It is clear from these figures that the unevolved functions are all
sharply peaked at high $z$ and get spread out and smoothed out with
evolution. The double peak structure noted for the direct production of
the ground state heavy baryons in our previous work~\cite{adamov1,adamov2}
has been moderated by the contributions from the excited states. The bound
$f_1 + g_1 \geq 2|h_1|$ is satisfied~\cite{soffer} and nearly saturated,
as expected in the heavy quark limit. 

The overall rate for a heavy baryon state to be a fragment of the
corresponding heavy quark is obtained by
integrating $f_1(z)$ over $z$. The results, which include the excited
intermediate states are tabulated here along with experimental data.
\begin{eqnarray}
 & Total \ Fragmentation \ Probabilities&   \nonumber \\
Particle& Experiment & Prediction  \nonumber \\
\Lambda_c & 5.6\pm 2.6\% [OPAL~\cite{opal}]& 3.9\%  \nonumber \\
\Xi_c     &                      & 0.59\% \nonumber \\
\Lambda_b & 7.6\pm 4.2\% [ALEPH~\cite{aleph}]&6.7\%  \nonumber 
\label{tb:tot}
\end{eqnarray}
These results are consistent with experiment within errors. The total
fragmentation probability for $\Xi_c$ is not given, although the
fractional rates for different states are known~\cite{CLEO} and were
stated in our previous work~\cite{adamov2}.

The predicted overall net polarization transfer from quark
to baryon is the integral of $g_1(z)/f_1(z)$. For these we obtain
$0.288, 0.239 \ {\rm and}\ 0.302$ for $\Lambda_c, \Xi_c\ {\rm and}\
\Lambda_b$,
respectively. The averaged value of the polarization measurements for
$\Lambda_b$ produced at LEP~\cite{newlep} are $-0.45\pm{0.18}$ compared to
the Standard Model expectation of $-0.94$ for the b-quark polarization. So
the value of net polarization transfer for $b\rightarrow\Lambda_b$ is 
$0.48\pm{0.19}$, consistent with our prediction of $0.30$. Note that
our prediction is based on integrating over all $z$, whereas the
experiment necessarily emphasizes a narrower range of $z$ centered on the
peak production region and there is some fluctuation of the value of
$g_1(z)/f_1(z)$ in that region.

\section{Discussion of the results.}

It is known that heavy quark fragmentation into {\it excited} charmonium
and bottomonium states contributes significantly to the high energy
hadronic production of $J/\psi, \ \psi' \ {\rm and} \ \Upsilon$ 
states~\cite{cdfonium}. The excited intermediate states do not make up
completely for the overall theoretical underestimate of the cross sections
for heavy quarkonium production (color octet~\cite{braaten_new} schemes or
alternatives are needed), but they are quite significant.  The situation
in the production of heavy baryons is less clear at this time. However,
the observation of the depolarization of the heavy quark provides an
opportunity to pin down the contribution of excited states and to compare
the theoretical model predictions based on QCD with the
experimental data. Such an effect for the $\Lambda _b$ was first
estimated by Falk and Peskin~\cite{peskin}. The heavy c and b baryons fall  
into their second or third category. The decay of the excited $\Sigma _b$,
$\Sigma _c$, $\Xi _b$ or $\Xi _c$ baryons is primarily due to pion
emission (with the exception of the radiative decay of $\Xi_c(2574)$
as we discussed above). The experimental data on the decay widths are
either unavailable or come with large margins of error. They can be
theoretically estimated for the heavy baryons~\cite{Yan}. The results
clearly indicate that the decay of the excited heavy baryons should
exhibit at least partial depolarization.

Such depolarization is more important in the light of the results
obtained in the direct fragmentation model~\cite{adamov1,adamov2}. Since
the number of the excited states that can contribute is large, they may
very well pollute the final sample of the ground
state baryons and result in an overall depolarization. Based on general,
model independent reasoning Falk and Peskin~\cite{peskin} estimated the
amount of depolarization: 
\begin{equation}
\frac{\Lambda _b(-\frac 12)}{\Lambda _b(+\frac 12)}=\frac{2(2-w_1)A}{%
9+A(5+2w_1)}
\label{eq:lambdaratio}
\end{equation}
with all heavy quarks initially taken to be in $+\frac 12$ state. The
parameter $A$ is the probability of producing the vector diquark (versus the
scalar); $w_1$ is the probability that the vector is in the helicity +1 or -
1 state. Note that when direct production of $\Lambda _b$ is much bigger
than the direct production of the excited states, as naively may be 
assumed to be the case, the depolarization is minimal. On the other hand,
the signature of our direct quark-diquark fragmentation model is the high
ratio of the excited state production, which leads to a sizeable
depolarization.

The original investigation of excited state contributions by Falk and
Peskin~\cite{peskin} was made without any dynamical details. What we have
done here is to go a step further, with an eye toward future high
statistics experiments, and to look at the dynamics of the
indirect fragmentation. We used a perturbative
calculation of excited baryon fragmentation and decay. The idea was simply
to incorporate pion decay into the direct quark-diquark model. The basic  
premise of the model still involves a heavy fast quark originating
from any high energy source, such as the $Z^0$ decay. Following that the
quark shakes off the gluon that breaks into a diquark antidiquark pair. If
the diquark has velocity similar to that of the heavy quark they fragment
into the baryon. The diquark could be either scalar, and then the baryon
produced is in the ground state (i.e. $\Lambda _b$), or vector, so that
the baryon is in one of the spin excited states.

Now note that in Falk and Peskin the final net polarization of
$\Lambda_Q $ (including $\frac 12$ and $\frac 32$ excited state
contributions), obtained from Eq.~\ref{eq:lambdaratio} is :
\begin{equation}
P=\frac{1+(1+4w_1)A/9}{1+A}
\end{equation}
That corresponds to the integral of $\frac{f_1}{g_1}$ over all values of
$z$. Obviously that integrated value is not enough to pinpoint both of
the parameters ($A$ and $w$). However, the net polarization of $\Lambda_Q$
produced purely via decays of excited states (direct fragmentation
excluded) is:
\begin{equation}
P=\frac{4w_1+1}9.
\end{equation}
Using this expression to obtain $w_1$ we find:

For $\Lambda _b$: $w_1=0.41$. Total polarization $P=0.295$

For $\Lambda _c$: $w1=0.39$. Total polarization $P=0.285$

In our previous paper we had $w_1=0.46$ for $\Lambda _b$ using the
direct fragmentation function of $\Lambda _b$ and $\Sigma _b$. The results
here are very close to this value although we used different
techniques to obtain it. Also,
these results assume that $w$ and $A$ are independent of $z$, which
generally is not correct (production rates may depend on the energies of the
gluon and that can be translated into $z$ dependence).

The parameter $A$ has not changed much from the last paper.

For $\Lambda _b$: $A=6$.

For $\Lambda _c$: $A=6.3$.

The various fragmentation functions that have been obtained in our model
have yet to be tested experimentally. The important features of those
functions that should appear in the data include their overall
normalizations, the resulting longitudianl polarizations, the ratios of
the three spin dependent functions and the characteristic double hump
dependence on $z$. The functional dependence on $z$ distinguishes this
model from the qualitative parameterization of the Peterson model. The
departure from a single sharp peak structure is a measure of the departure
from the heavy quark limit. So careful measurements of the $z$ dependence
of any of the functions, $f_1(z), g_1(z)\ {\rm or}\ h_1(z)$ will be very
revealing.


\appendix
\section*{Appendix}
In this appendix we list all relevant scalar products of two 4-vectors  
beginning with all combinations not involving spin. In each case we show
the limiting value as the fragmenting quark's energy $q_0$ and 3-momentum
$\left| \overrightarrow{q}\right|$ become large.

a) $(q\overline{q})$

\[
(q\overline{q})=q_0\overline{q}_0+\left| \overrightarrow{q}\right| ^2=\sqrt{%
m_q^2+\left| \overrightarrow{q}\right| ^2}\sqrt{s_q+\left| \overrightarrow{q}%
\right| ^2}+\left| \overrightarrow{q}\right| ^2\rightarrow 2\left| 
\overrightarrow{q}\right| ^2
\]

b) $(p\overline{q})$%
\[
(p\overline{q})=p_0q_0+p_l\left| \overrightarrow{q}\right| \rightarrow
2p_l\left| \overrightarrow{q}\right| =2x_1\left| \overrightarrow{q}\right|
^2
\]

c) $(l\overline{q})$%
\[
(l\overline{q})\rightarrow 2z\left| \overrightarrow{q}\right| ^2
\]

d) $(n\overline{q})$%
\[
(n\overline{q})=q_0-\left| \overrightarrow{q}\right| =\sqrt{m_q^2+\left| 
\overrightarrow{q}\right| ^2}-\left| \overrightarrow{q}\right| \rightarrow 
\frac{m_q^2}{2\left| \overrightarrow{q}\right| }
\]

e) $(qp)$ 

Note that the intermediate quark (momentum $q - k$) is taken
as on-shell and the mass of the hadron (momentum $p$) $M_{\Sigma}$ is
approximately $m_{diquark} + m_{quark}$. Hence $q = k +
\frac{m_q}{M_{\Sigma}}p\ {\rm and}\ p^{\prime} = k -
\frac{m_d}{M_{\Sigma}}$. The quantity $s_q = q^2$. Hence:

\[
(qp)=m_qM_\Sigma +(kp)
\]

\[
(qq)=m_q^2+2\frac{m_q}{M_\Sigma }(kp)+k^2=s_q 
\]

\[
k^2=2\frac{m_d}{M_\Sigma }(kp) 
\]
So finally
\[
(qp)=m_qM_\Sigma +\frac{(s-m_q^2)}2.
\]

f) ($nq)$%
\[
(nq)=q_0+\left| \overrightarrow{q}\right| \rightarrow 2\left| 
\overrightarrow{q}\right| 
\]

g) $(lp)$%
\[
p=l+\overline{\pi }
\]
\[
(p-l)^2=m_\pi ^2
\]
\[
M_\Sigma ^2+M_\Lambda ^2-2(pl)=m_\pi ^2
\]
\[
(lp)=\frac{M_\Sigma ^2+M_\Lambda ^2-m_\pi ^2}2
\]

h) $(ql)$

Evaluating this expression requires careful definitions of coordinate
systems. Let $\hat{z} = \hat{q}\ {\rm and}\ \hat{x} = \vec{S}_{T}(q)$
define
the X-Z plane for the incoming quark (with transverse spin vector
also called $s^{qt}$). Let
($\theta, \phi$) be the polar
and azimuthal angles for $\vec{p}$. Define a second
frame of reference in which $\hat{z}^{\prime} = \hat{p}$  is the
polar axis and $\hat{x}^{\prime} =
[\hat{x}cos(\theta)-\hat{z}sin(\theta)cos(\phi)]/ 
[1-sin^2(\theta)sin^2(\phi)]^{\frac{1}{2}}$ so that $\hat{x}^{\prime}$ is 
defined by the excited baryon's transverse spin direction.
We can express $l$ in the primed coordinates as
$l=(l_0,l_{pt}\cos (\varphi ),l_{pt}\sin (\varphi ),l_{pl})$. The azimuth
of $\vec{q}$ in the primed coordinates is approximately $\pi -\phi $. It
is more precisely $=\pi -\phi +O(\frac 1{\left| \vec{p}\right| })$ in the
relevant large momentum limit. Using
that relation we get:   
\[
(\overrightarrow{q}\cdot \overrightarrow{l})\rightarrow \left| 
\overrightarrow{q}\right| \frac{p_{ql}}{\left| \overrightarrow{p}\right| }%
l_{pl}+\frac{\left| \overrightarrow{q}\right| }{\left| \overrightarrow{p}%
\right| }p_{qt}l_{pt}\cos (\phi ^{\prime }-\varphi )
\]

Then after several steps the appropriate limiting value is given by

\[
(ql)\rightarrow -{\frac {x_2}{x_1}}M_\Sigma
^2+(qp)x_2+\frac{(lp)}{x_1}-\frac{p_{qt}l_{pt}}%
{x_1}\cos (\phi +\varphi )
\]

i) $(np)$%
\[
(np)=p_0+p_{ql}\rightarrow 2x_1\left| \overrightarrow{q}\right| 
\]

j) $(nl)$%
\[
(nl)=l_0+l_{ql}\rightarrow 2z\left| \overrightarrow{q}\right| 
\]

These complete all scalar products not involving the spin vectors. Now
we consider transverse spin vectors also. We only have two distinct
ones: $s^{qt}$ and $s^{lt}$.

k) $(s^{qt}p)$%
\[
(s^{qt}p)=-p_{qt}\cos (\phi )
\]

l) $(s^{qt}l)$

This also requires going into the primed frame of $p$. In that frame:

\[
(s^{qt}p)=\left( 0,(1+O(\frac 1{\left| \overrightarrow{p}\right|
})),O(\frac
1{\left| \overrightarrow{p}\right| }),(\frac{p_{qt}\cos (\phi )}{%
\overrightarrow{p}}+O(\frac 1{\left| \overrightarrow{p}\right| ^2}))\right) 
\]

\[
(s^{qt}l)\rightarrow -l_{pt}\cos (\varphi )-\frac{l_{pl}}{\left| 
\overrightarrow{p}\right| }p_{qt}\cos (\phi )+O(\frac 1{\left| 
\overrightarrow{p}\right| })
\]
\[
(s^{qt}l)\rightarrow -l_{pt}\cos (\varphi )-\frac{p_{qt}\cos (\phi )}{x_2}
\]

m) $(s^{qt}n)$%
\[
(s^{qt}n)\rightarrow \frac{p_{qt}\cos (\phi )}{\left|\overrightarrow{p}\right|}
\]

n) $(s^{lt}\overline{q})$

In the primed frame:
\[
(s^{lt}p)=\left( 0,(1+O(\frac 1{\left| \overrightarrow{p}\right|
})),O(\frac
1{\left| \overrightarrow{p}\right| }),(-\frac{p_{qt}\cos (\phi )}{%
\overrightarrow{p}}+O(\frac 1{\left| \overrightarrow{p}\right| ^2}))\right) 
\]
\[
(s^{lt}\overline{q})\rightarrow -\frac{q_{pl}}{\left| \overrightarrow{p}%
\right| }p_{qt}\cos (\phi )
\]
\[
(s^{lt}\overline{q})\rightarrow -\frac 1x_1p_{qt}\cos (\phi )
\]

o) $(s^{lt}q)$

\[
(s^{lt}\overline{q})\rightarrow \frac 1x_1p_{qt}\cos (\phi )
\]

p) $(s^{lt}p)$%
\[
(s^{lt}p)\rightarrow p_{qt}\cos (\phi )
\]


\begin{thebibliography}{99}
\bibitem{isgur}  N. Isgur and M.B. Wise, \Journal{\PLB}{232}{113}{1989}.

\bibitem{peterson}  C. Peterson, {\it et
al.},\Journal{\PRD}{27}{105}{1983}; R.D. Field and R.P. Feynman,
\Journal{NPB}{136}{1}{1978}.

\bibitem{hoodbhoy}  M.Nzar and P.Hoodbhoy, \Journal{\PRD}{51}{32}{1995}.

\bibitem{lund}  B. Andersson, {\it et al.}, {Phys. Reports}{\bf 97}, 31
(1983); T. Sj\"{o}strand, {Comput. Phys. Commun.}{\bf 39}, 347 (1986); G.
Marchesini and B.R. Webber, \Journal{\NPB}{238}{1}{1984}.

\bibitem{ji}  R.L. Jaffe and X. Ji, \Journal{\PRL}{71}{2547}{1993}.

\bibitem{randall}  R. L. Jaffe and L. Randall,
\Journal{\NPB}{412}{79}{1994}.

\bibitem{chang}  C.-H. Chang and Y.-Q. Chen,
\Journal{\PLB}{284}{127}{1992}.

\bibitem{braaten}E. Braaten, K. Cheung, S. Fleming, T.C. Yuan,
\Journal{\PRD}{51}{4819}{1995}; and references contained therein.

\bibitem{adamov1}  A. Adamov and G. R. Goldstein, in {\it Diquarks III},
editors M. Anselmino and E. Predazzi (World Scientific, Singapore 1998)
p.218.

\bibitem{adamov2}  A. Adamov and G. R. Goldstein,
\Journal{\PRD}{56}{7381}{1997}.

\bibitem{russians}  A.P. Marteynenko and V.A. Saleev,
\Journal{\PLB}{385}{297}{1996}.

\bibitem{russians2}  V.A. Saleev, \Journal{\PLB}{426}{384}{1998};
{\it ibid}, \Journal{\MPLA}{14}{2615}{1999}.

\bibitem{mulders} R. Jakob, P.J. Mulders and J. Rodrigues,
\Journal{\NPA}{626}{937}{1997}.

\bibitem{peskin}  A.F. Falk and M.E. Peskin,
\Journal{\PRD}{49}{3320}{1994}.

\bibitem{kwong} W. Kwong, J. Rosner, and C. Quigg,
Annu. Rev. Nucl. Sci. {\bf 37}, 325 (1987).

\bibitem{gold1}G. R. Goldstein, J. Maharana,
\Journal{\NCA}{59}{393}{1980}; G. R. Goldstein in {\it Diquarks}, editors
M. Anselmino and E. Predazzi (World Scientific, Singapore 1989) p. 159;
H. Liebl and G. R. Goldstein, \Journal{\PLB}{343}{363}{1995}.

\bibitem{transversity}  G.R. Goldstein and M.J. Moravcsik, {Ann. Phys. (NY)}%
{\bf 98}, 128 (1976); {Ann. Phys. (NY)}{\bf 142}, 219 (1982); {Ann. Phys.
(NY)}{\bf 195}, 213 (1989).

\bibitem{soffer} J. Soffer, \Journal{\PRL}{74}{1292}{1995}; 
G. R. Goldstein, R. L. Jaffe, X. Ji, \Journal{\PRD}{52}{5006}{1995}.

\bibitem{quigg} E.J. Eichten and C. Quigg, \Journal{\PRD}{52}{1726}{1995}.

\bibitem{artru} X. Artru and M. Mekhfi, \Journal{\ZPA}{45}{669}{1990}.

\bibitem{cleo1} P. Avery, {\it et al.} (CLEO
Colaboration) \Journal{\PRD}{43}{3599}{1991}.

\bibitem{opal}  G. Alexander, {\it et al.}, \Journal{\ZPC}{72}{1}{1996}.

\bibitem{aleph}  U. Becker, {\it et al.} (ALEPH Collaboration), preprint
hep-ex/9608004 (1996).

\bibitem{CLEO}L. Gibbons, {\it et al.}, \Journal{\PRL}{77}{810}{1996};
P. Avery, {\it et al.}, \Journal{\PRL}{75}{4364}{1995};
K.W. Edwards, {\it et al.}, \Journal{\PLB}{373}{261}{1996}.

\bibitem{newlep} ALEPH, {\it et al.}, preprint SLAC-PUB-8492 (2000).

\bibitem{cdfonium} T. Affolder, {\it et al.} (CDF Collaboration)
\Journal{\PRL}{84}{2094}{2000}.

\bibitem{braaten_new} E. Braaten, S. Fleming and A.K. Leibovich, preprint
hep-ph/0008091 (2000), and references contained therein.

\bibitem{Yan}  Tung-Mow Yan, {\it et al.}, \Journal{\PRD}{46}{1148}{1992}.

\end{thebibliography}
\end{document}